\documentclass[12pt,usenames,dvipsnames]{article}

\usepackage{latexsym}
\usepackage{amssymb,amsfonts,amsmath}
\usepackage{graphicx} 
\usepackage{indentfirst}
\usepackage{bbm}
\usepackage{amssymb}
\usepackage{verbatim}
\usepackage{amsmath, amsthm,amssymb}
\usepackage{mathrsfs}
\usepackage{hyperref}
\usepackage{amsfonts}
\usepackage{dsfont}
\usepackage{cite}
\usepackage{xcolor}
\usepackage{enumerate}
\usepackage{cleveref}

\parskip=\medskipamount

\newcommand {\cD}{{\cal D}}
\newcommand {\cE}{{\cal E}}

\newcommand {\cH}{{\cal H}}

\newcommand {\cN}{{\cal N}}
\newcommand {\cO}{{\cal O}}

\def\a{\alpha}

\def\b{\beta}
\def\c{\chi}
\def\d{\delta}

\def\f{\phi}
\def\g{\gamma}
\def\G{\Gamma}

\def\k{\kappa}
\def\l{\lambda}

\def\n{\nu}
\def\o{\omega}

\def\q{\theta}

\def\s{\sigma}
\def\t{\tau}

\def\x{\xi}

\def\D{\Delta}
\def\F{\Phi}
\def\J{\Psi}
\def\L{\Lambda}
\def\O{\Omega}

\def\S{\Sigma}
\def\U{\Upsilon}
\def\X{\Xi}

\def\rd{{\rm d}}
\def\ri{{\rm i}}
\def\re{{\rm e}}

\newcommand{\ad}{{\dot{\alpha}}}                           %
\newcommand{\bd}{{\dot{\beta}}}                            %
\newcommand{\ve}{\varepsilon}                            %

\newcommand{\pa}{\partial}                           %
\newcommand{\hf}{\frac12}

\newcommand{\vf}{\varphi}

\newcommand{\be}{\begin{equation}}
\newcommand{\ee}{\end{equation}}
\newcommand{\bea}{\begin{eqnarray}}
\newcommand{\eea}{\end{eqnarray}}
\newcommand{\non}{\nonumber}

\newcommand{\bm}[1]{\mbox{\boldmath$#1$}}

\def\double #1{#1{\hbox{\kern-2pt $#1$}}}

\newif\ifdtup

\newcommand{\bsubeq}{\begin{subequations}}
\newcommand{\esubeq}{\end{subequations}}

\numberwithin{equation}{section}

\newcommand{\sSL}{\mathsf{SL}}

\newcommand{\sU}{\mathsf{U}}

\begin{document}

\begin{titlepage}
\begin{flushright}
April, 2024 \\
\end{flushright}
\vspace{5mm}

\begin{center}
{\Large \bf 
Higher-derivative deformations of the ModMax theory
}
\end{center}

\begin{center}

{\bf Sergei M. Kuzenko and Emmanouil S. N. Raptakis} \\
\vspace{5mm}

\footnotesize{
{\it Department of Physics M013, The University of Western Australia\\
35 Stirling Highway, Perth W.A. 6009, Australia}}  
~\\
\vspace{2mm}
~\\
Email: \texttt{ 
sergei.kuzenko@uwa.edu.au, emmanouil.raptakis@uwa.edu.au}\\
\vspace{2mm}

\end{center}

\begin{abstract}
\baselineskip=14pt
We present higher-derivative deformations of the ModMax theory which preserve both $\mathsf{U}(1)$ duality symmetry and Weyl invariance. In particular, we single out a class of deformations expected to describe a low-energy effective action for the ModMax theory. We also elaborate on (higher-derivative) deformations of the ${\cal N}=1$ super ModMax theory. 
\end{abstract}
\vspace{5mm}

\vfill

\vfill
\end{titlepage}

\newpage
\renewcommand{\thefootnote}{\arabic{footnote}}
\setcounter{footnote}{0}

\allowdisplaybreaks

\section{Introduction}

Recently much interest has been attracted to the 
unique conformal $\sU(1)$ duality-invariant nonlinear electrodynamics proposed 
by Bandos, Lechner, Sorokin and Townsend \cite{BLST}. It is described by the Lagrangian 
\bea
\label{2.7}
L_{\rm MM}(F_{ab}) 
=L_{\rm MM}(\o, \bar \o) = - \hf  \big( {\o} + {\bar \o}\big)\cosh \g 
+   {\sqrt{\o\bar \o} } \sinh \g ~, \qquad 
\eea
 where $\g $ is a non-negative real coupling constant,\footnote{The parameter $\g$ is restricted to be non-negative since superluminal propagation is possible for $\g<0$ \cite{BLST}.} and  $\o$ is defined by 
\bea
\o = \a + {\rm i} \, \b~, \qquad \a = \frac{1}{4} \, F^{ab} F_{ab}~, \quad 
\b = \frac{1}{4} \, \widetilde{F}^{ab} {F}_{ab} ~,
\label{EMinvariants}
\eea
with $\widetilde{F}^{ab} = \hf \ve^{abcd} F_{cd}$ the Hodge dual of the field strength $F_{ab}$.
This model for nonlinear electrodynamics was called the ModMax theory, that is  modified Maxwell electrodynamics.
Originally it was derived using the Hamiltonian methods for nonlinear electrodynamics advocated by Bialynicki-Birula \cite{B-B}. Ten days after \cite{BLST} had been released in the arXiv, the theory was re-derived in \cite{Kosyakov} using the  
Gaillard-Zumino-Gibbons-Rasheed (GZGR) formalism of duality rotations for nonlinear electrodynamics \cite{GZ1,GR1,GR2,GZ2,GZ3}. The ModMax theory was also shown to be a unique conformal solution \cite{K21} within the Ivanov-Zupnik (IZ) approach to $\sU(1)$ duality-invariant nonlinear electrodynamics \cite{IZ_N3,IZ1,IZ2}.

There is an interesting historical curiosity concerning the ModMax theory. In the late 1990s and early 2000s, much work was done to describe general $\sU(1)$ duality-invariant Lagrangians for nonlinear electrodynamics \cite{GR1,GZ2,GZ3,Hatsuda:1999ys,IZ1,IZ2} 
of which the most prominent example is the Born-Infeld theory \cite{Born:1934gh}
\begin{align}
L_{\rm BI} = \frac{1}{g^2} \left\{
1 - \sqrt{- \det (\eta_{ab} + g F_{ab} )} 
\right\} 
=
  \frac{1}{g^2} \left\{ 
1 - \sqrt{1 + g^2 (\o + \bar \o )
+{1 \over 4}g^4 (\o - \bar \o )^2 } 
\right\}~, ~~
\end{align}
with $g$ the coupling constant. A natural question is the following: Why wasn't the ModMax theory discovered at that time? One of the reasons is that all self-dual models for nonlinear electrodynamics which were  studied before \cite{BLST} were assumed to possess a well defined weak-field limit, and the existence of such a limit is an intrinsic feature of low-energy effective actions. As is clear from \eqref{2.7}, the ModMax theory does not have a weak-field limit.\footnote{The $\cN=1$ supersymmetric extension of the ModMax theory \cite{K21,Bandos:2021rqy} also does not have a weak-field limit. This is in contrast with the superconformal $\sU(1)$ duality-invariant model for the $\cN=2$ vector multiplet proposed in \cite{KT1,K21}. The latter theory is described in terms of the  $\cN=2$ {\it reduced chiral} superfield strength, $W(x,\q) = \vf(x) + \q^\a_i \q^{\b i} F_{(\a\b)} (x) +\dots$, and its conjugate. The action  functional is not analytic with respect to the physical scalars $\vf$ and $\bar \vf$, which are required to possess non-zero VEVs, but it is analytic in the electromagnetic field strength $F_{\a \b}$ and  and its conjugate $\bar F_{\ad\bd}$ related to $F_{ab}$ according to \eqref{211}.
}
To make this point clearer, we remind the reader that every $\sU(1)$ duality-invariant 
nonlinear electrodynamics is described by a 
 Lorentz-invariant Lagrangian $L(F_{ab})$ which is a solution 
to the self-duality equation \cite{B-B,GR1,GZ2}
\bea
\widetilde{G}^{ab}G_{ab}  +  \widetilde{F}^{ab}F_{ab} = 0~,
\label{SDequation}
\eea
where
\bea
\widetilde{G}^{ab} (F):=
\hf \, \ve^{abcd}\, G_{cd}(F) =
2 \, \frac{\pa L(F)}{\pa F_{ab}}~.
\eea
The self-duality equation guarantees invariance under $\sU(1)$ duality rotations
\bea
\d F_{ab} = \vf G_{ab}~, \qquad \d G_{ab} = -\vf F_{ab}~,
\eea
on the mass shell.
If the Lagrangian $L(F_{ab})$ is expressed in terms of the electromagnetic invariants $\o$ and $\bar \o$, eq. \eqref{EMinvariants}, 
then the self-duality equation \eqref{SDequation} turns into
\bea
{\rm Im}\, \bigg\{ \frac{\pa (\o \, \L) }{\pa \o}
- \bar{\o}\,
\left( \frac{\pa (\o \, \L )  }{\pa \o} \right)^2 \bigg\} = 0~,
\label{GZ4}
\eea
where $\L(\o,\bar \o)$ is a real function related to the Lagrangian $L(\o, \bar \o)$ by the rule 
\bea
L(\o, \bar \o)  = -\hf \, \Big( \o + \bar{\o} \Big) +
\o \, \bar{\o} \; \L (\o, \bar{\o} )~,
\eea
see \cite{KT2} for the technical details.\footnote{For any $\sU(1)$ duality-invariant model for nonlinear electrodynamics, 
its  compact duality group $\sU(1)$
can be enhanced to the non-compact $\sSL(2,{\mathbb R})$ group by coupling the electromagnetic field to the dilaton  and axion fields
\cite{GR2,GZ2,GZ3}. In the case of the ModMax theory, this was discussed in \cite{K21, Babaei-Aghbolagh:2022itg}. It is also worth pointing out a Galilean cousin of the ModMax theory introduced in \cite{Banerjee:2022sza}; it is invariant under Galilean conformal transformations.}
For the nonlinear theory to possess a weak-field limit, the self-coupling $\L(\o,\bar \o)$ must be a real analytic function in a neighbourhood of $\o=0$, 
\bea
\L (\o , \bar \o ) ~=~ \sum_{n=0}^{\infty} ~
\sum_{p+q =n} c_{p,q} \; \o^p {\bar \o}^q~,
\qquad \quad c_{p,q}=c_{q,p} ~\in~ {\Bbb R}~.
\label{self-coupling}
\eea
Here the reality of the Taylor coefficients follows from \eqref{GZ4}, see \cite{KT2} for more comments.
However, the ModMax theory \eqref{2.7} is characterised by the function \cite{K21}
\bea
\L_{\rm MM}(\o, \bar \o) = \frac{\sinh \g }{\sqrt{\o\bar \o} } 
- \hf  \Big( \frac{1}{\o} + \frac{1}{\bar \o}\Big)(\cosh \g -1)
~,
\label{LambdaMM}
\eea
which is evidently not of the form \eqref{self-coupling}. 

In accordance with the above discussion, it appears that the ModMax theory cannot arise as a perturbative low-energy effective action. However, its origin as a non-perturbative quantum correction is not excluded. It is also natural to wonder whether a quantum version of the ModMax theory may be defined and, if so, what is the explicit structure of the corresponding counterterms, Weyl anomalies etc.? While this theory clearly does not possess a Poincar\'e-invariant vacuum state, one may nevertheless still try to carry out a path-integral analysis. One-loop logarithmic divergences should respect the Weyl and duality symmetries (related formal arguments may be found in \cite{FT,RT}), and the only possible functional structure of the form $\int \rd^4x \, e\, \mathfrak{L} (F_{ab})$ is given by \eqref{2.7}. Such quantum corrections are absent \cite{Pinelli}.\footnote{Similar results hold for chiral bosons in two dimensions \cite{Ebert:2024zwv}.} 
Higher-derivative quantum corrections are possible, and thus it becomes important to address the problem of finding consistent higher-derivative deformations of the ModMax theory.   

A generalisation of the GZGR formalism to the case of $\sU(1)$ duality-invariant theories with higher derivatives
was sketched in \cite{KT2}.\footnote{The complete formalism applicable to higher-derivative systems was given in \cite{KT1,KT2} for $\cN=1$ and $\cN=2$ supersymmetric duality-invariant theories.}
Two modifications are required. 
Firstly, the definition of $\widetilde G$ is replaced with
\bea
\widetilde G^{ab}[F] =2 \,{\delta S[F] \over\delta F_{ab}}~.
\label{1.3}
\eea
Secondly, the self-duality equation \eqref{SDequation} is replaced with 
\bea
\int \rd^4 x\, e\,  \left( \widetilde{G}^{ab}G_{ab}  +  \widetilde{F}^{ab}F_{ab} \right)= 0~.
\label{1.4}
\eea
It is assumed in \eqref{1.3} and \eqref{1.4} that $S[F]$ is unambiguously defined 
as a functional of an unconstrained two-form $F_{ab}$, i.e. no dependence 
on $\pa_b \widetilde{F}^{ab} $ is present. In equation \eqref{1.4}, $F_{ab}$ 
is considered to be an unconstrained bivector. 
Duality-invariant theories with higher derivatives naturally occur in 
$\cN=2$ supersymmetry \cite{KT2, KT1}.
Further aspects of duality-invariant theories 
with higher derivatives were studied in, e.g., \cite{AFZ,Chemissany:2011yv,AF,AFT}.

Within the GZGR approach or its supersymmetric extensions, a consistent nonlinear deformation of a given $\sU(1)$ duality-invariant theory (say, Maxwell's theory) is typically derived in perturbation theory and requires an infinite number of terms for the deformed action to satisfy the self-duality equation \eqref{SDequation} or its supersymmetric extensions \cite{KT1,KT2, Chemissany:2011yv, Carrasco:2011jv, Broedel:2012gf}. Hence, a closed-form expression for the deformed action is difficult to obtain in such a setting. However, within the IZ approach, which makes use of auxiliary variables $V_{ab} =-V_{ba}$, all information about  the given duality-invariant theory is encoded in its $\sU(1)$-invariant interaction Lagrangian $\mathfrak{L}^{\rm int} (V_{ab})$. Its $\sU(1)$-invariance is equivalent to the self-duality equation \eqref{SDequation}.  A consistent deformation of the theory amounts to deforming  $\mathfrak{L}^{\rm int} (V_{ab}) \to  \hat{\mathfrak{L}}^{\rm int} (V_{ab})$ in such a way that $\hat{\mathfrak{L}}^{\rm int} (V_{ab})$ is also $\sU(1)$ invariant. In the presence of the auxiliary variables, the deformed theory is given in closed form. The IZ approach has also been extended to $\sU(1)$ duality-invariant theories with higher derivatives \cite{Ivanov:2012bq, ILZ2}.\footnote{It should be pointed out that Ref. \cite{BN} and later \cite{Carrasco:2011jv,Chemissany:2011yv} advocated the so-called twisted self-duality constraint
as a systematic procedure to generate duality-invariant theories. 
However, it was demonstrated \cite{Ivanov:2012bq} that the construction of  
\cite{BN, Carrasco:2011jv,Chemissany:2011yv} naturally originates within the IZ approach proposed a decade earlier.
Specifically, the twisted self-duality constraint 
corresponds to an equation of motion in the approach of \cite{IZ1,IZ2}.}

This paper is organised as follows. In section \ref{Section2} we describe several families of higher-derivative deformations of the ModMax electrodynamics, including special classes expected to appear in loop quantum corrections to the theory. Such deformations are determined within the auxiliary variable formulation of \cite{IZ1,IZ2,IZ_N3}. Thus, in section \ref{Section3} we consider two important models and eliminate such variables in perturbation theory. Section \ref{Section4} is devoted to higher-derivative deformations of the $\cN=1$ super ModMax theory as an extension of the non-supersymmetric analysis of section \ref{Section2}. Concluding comments are provided in section \ref{Section5}.

\section{Higher-derivative deformations of ModMax}
\label{Section2}

A natural framework to generate $\sU(1)$ duality-invariant models for nonlinear electrodynamics is the IZ approach \cite{IZ_N3,IZ1,IZ2}.
In the case of theories without higher derivatives, it is a reformulation of the GZGR formalism  \cite{GZ1,GR1,GR2,GZ2,GZ3} which is obtained by replacing $L(F_{ab}) \to \mathfrak{L}(F_{ab} , V_{ab})$, where  $V_{ab}=-V_{ba}$ is an auxiliary unconstrained bivector. The latter is equivalent to a pair of symmetric rank-2 spinors, $V_{\a\b}= V_{\b\a}$ and its conjugate $\bar  V_{\ad\bd}$, which are defined by 
\bea
(\s^a)_{\a \ad} (\s^b)_{\b \bd} V_{ab} = 2 \ve_{\a \b} \,
{\bar V}_{\ad \bd} + 2 \ve_{\ad \bd} \, V_{\a \b}~. 
\label{211}
\eea
Our two-component spinor notation and conventions, including the definition of the relativistic Pauli matrices $(\s_a)_{\a\ad}$, follow \cite{WB,Buchbinder:1998qv,Kuzenko:2022skv}. 

The new Lagrangian $\mathfrak{L}$ is at most quadratic in
the electromagnetic field strength $F_{ab}$, while the self-interaction is described 
by a nonlinear function of the auxiliary variables, $\mathfrak{L}^{\rm int} (V_{ab})$,
\bea
\label{2.4}
 \mathfrak{L}(F_{ab} , V_{ab}) = \frac{1}{4} F^{ab}F_{ab} +\hf V^{ab}V_{ab} 
 - V^{ab}F_{ab} + \mathfrak{L}^{\rm int} (V_{ab})~.
\eea 
The original theory  $L(F_{ab})$ is derived from 
$\mathfrak{L}(F_{ab} , V_{ab})$ by integrating out the auxiliary variables using their algebraic equations of motion.
In terms of $\mathfrak{L}(F_{ab} , V_{ab})$, the condition of $\sU(1)$ duality invariance 
was shown \cite{IZ1,IZ2}  to be equivalent to the requirement that the self-interaction
\bea
\mathfrak{L}^{\rm int} (V_{ab}) = \mathfrak{L}^{\rm int} (\n, \bar \n)~, \qquad \n:=V^{\a\b}V_{\a\b}
\eea
be invariant under linear $\sU(1)$  transformations $\n \to \re^{2 \ri \vf} \n$, with $\vf \in \mathbb R$, therefore
\bea
\mathfrak{L}^{\rm int} (\n, \bar \n)= f (\n \bar \n)~.
\eea
The ModMax theory corresponds to the choice \cite{K21}
\bea
\mathfrak{L}^{\rm int}_{\rm MM} = \k \sqrt{\n \bar \n}~, \qquad \sinh \g = \frac{\k}{1-(\k/2)^2} ~.
\label{MM2.4}
\eea

In the case of $\sU(1)$ duality-invariant theories with higher derivatives, 
the self-coupling $\mathfrak{L}^{\rm int} $ becomes a multivariable function, $\mathfrak{L}^{\rm int} (V_{ab}) \to 
\mathfrak{L}^{\rm int} (V_{ab}, \nabla_c V_{ab}, \nabla_c \nabla_d V_{ab}, \dots )$, and it becomes more economical to work with the action functional. The IZ reformulation is obtained 
by replacing the action functional $S[F]$ with a first-order action\footnote{Here $\mathfrak{S}^{\rm int} [V] = \int \rd^4 x \,e \, \mathfrak{L}^{\rm int} (V_{ab}, \nabla_c V_{ab}, \nabla_c \nabla_d V_{ab}, \dots )$.} 
\bea
\mathfrak{S}[F, V] = \int \rd^4 x \,e \,\left\{ \frac{1}{4} F^{ab}F_{ab} +\hf V^{ab}V_{ab} 
 - V^{ab}F_{ab} \right\}+ \mathfrak{S}^{\rm int} [V]
\label{AuxAction}
 \eea
such that imposing the equation of motion 
\bea
\frac{\d}{\d V_{ab} }  \mathfrak{S}[F, V] =0
\label{AuxEoM}
\eea
reduces the action \eqref{AuxAction} to $S[F]$. It may be shown that the self-duality equation \eqref{1.4} turns into\footnote{This condition has a natural generalisation to $4n$ dimensions \cite{Kuzenko:2019nlm}.} 
\bea
\int \rd^4 x \,e \, \widetilde{V}_{ab}  \frac{\d}{\d V_{ab} }  \mathfrak{S}^{\rm int}[ V] =0~.
\label{invariance1}
\eea
In the case that the interaction has the form $\mathfrak{S}^{\rm int}[ V] =  \mathfrak{S}^{\rm int}[ \n, \bar \n]$, \eqref{invariance1} is equivalent to the condition of manifest $\sU(1)$ invariance
\bea 
\mathfrak{S}^{\rm int}[ \re^{2\ri \vf} \n, \re^{-2\ri \vf}   \bar \n] = \mathfrak{S}^{\rm int}[ \n, \bar \n]~, \qquad \vf \in {\mathbb R}~.
\label{invariance2}
\eea

We are interested in those higher-derivative deformations of the ModMax theory which may contribute to a low-energy effective action of the theory. An important insight is obtained by 
considering the in-out vacuum amplitude for the ModMax theory 
\bea
Z = \int [\mathfrak{D} A_a] [\mathfrak{D} V_{ab} ]  \,\d\big[ \nabla_a A^a - \x\big ] {\rm Det}(\nabla^2 ) \,
\exp \left\{ \frac{\ri} {\hbar} \mathfrak{S}_{\rm MM} [F,V] \right\}~,
\label{in-out}
\eea
where $A_a$ is the gauge potential, 
$F_{ab} = \nabla_a A_b - \nabla_b A_a$, and $\x(x)$ is a background scalar field.\footnote{The  in-out vacuum amplitude is independent of $\x(x)$, in accordance with \cite{FP}.}   
In accordance with \eqref{2.4} and \eqref{MM2.4},
the functional $\hbar^{-1} \mathfrak{S}_{\rm MM}[F,V]$ is invariant under 
rescalings\footnote{Any rigid rescaling $A_a(x) \to \l A_a(x) $ is a symmetry of the ModMax theory \eqref{2.7} in the sense that it takes every solution of the equations of motion to another solution.}  
\bea
\hbar \to \l^2 \hbar ~, \qquad F_{ab}(x) \to \l F_{ab}(x)~, \qquad V_{ab}(x) \to \l V_{ab}(x)~.
\label{scale}
\eea
Formally, the effective action is expected to possess such a scale symmetry.  
Thus it is natural to assume that (a local part of) the effective action has the form
\begin{subequations} \label{LEEA}
\bea
\label{2.11a}
\G_{\rm MM} [F,V] = \mathfrak{S}_{\rm MM} [F,V] + \sum_{n=1}^{\infty} \hbar^n \G^{(n)} [V]
\eea
and possesses the following properties: 
 (i) $\hbar^{-1} \G_{\rm MM}[F,V]$ is 
invariant under \eqref{scale}; (ii) each functional $\G^{(n)} [V] $ is Weyl invariant;
and (iii) each functional $\G^{(n)} [V]$ obeys the condition \eqref{invariance1}.\footnote{These properties imply, in particular, that the ModMax coupling \eqref{MM2.4} cannot be generated as a loop quantum correction.} 
A solution to these requirements is given by 
\bea
\G^{(n)} [V] = g_n \int \rd^4 x \,e \, 
\frac{ \big[ \square_c (\n\bar \n)^{ 1/8} \big]^{2n} } {  (\n\bar \n)^{ (3n-2)/4}}~, 
\eea
\end{subequations}
where $g_n$ is a dimensionless numerical factor, and $\square_c:= (\nabla^2 - \frac 16 R)$ is the conformal d'Alembertian, with $\nabla^2 = \nabla^a \nabla_a$.\footnote{The operator $\square_c$ is conformal when acting on the space of primary dimension-one scalar fields.}

On the other hand, if we are only interested in 
a Weyl-invariant functional obeying  
 the condition \eqref{invariance1}, then a more general action is allowed
 \bea
\hat  \G_{\rm MM} [F,V] = \mathfrak{S}_{\rm MM} [F,V] + 
 \int \rd^4 x \,e \, \sqrt{\n \bar \n} \, 
\mathfrak{F} \left( \S
\right)~, \qquad \S:=\frac{ \square_c (\n\bar \n)^{ 1/8}  } {  (\n\bar \n)^{3/8}} ~,
\label{2.12}
\eea
for some real function $\mathfrak{F} (x) $ of a real argument. 

The specific feature of the functional \eqref{2.12}, including its special case \eqref{LEEA}, is that the auxiliary field appears in the deformation term only via the combination $\n \bar \n$. Another choice is to replace $\mathfrak F$ in \eqref{2.12} with a multivariable function of the form 
\bea
\mathfrak{H} ( \U, \bar \U, \Xi_n, \bar \X_n)~, \qquad \U := \frac{ \bar \n^{1/4} \square_c \n^{ 1/4}  } {  \sqrt{\n\bar \n}} ~,
\qquad \X_n:= \frac{ \bar \J^n \D_0 \J^n  } {  \sqrt{ \n\bar \n}} ~, \qquad  \J := \frac{\n}{\bar \n}~.
\label{2.13}
\eea
Here $\D_0$ denotes the Fradkin-Tseytlin operator \cite{FT1982}
\bea
\Delta_0 = (\nabla^a \nabla_a)^2 + 2 \nabla^a \big(
	 {R}_{ab} \,\nabla^b 
	- \tfrac{1}{3} {R} \,\nabla_a
	\big)~,
\label{1.5}
\eea
which is conformal when acting on the space of Weyl-neutral scalar fields.\footnote{This operator was  re-discovered by Paneitz in 1983 \cite{Paneitz} and Riegert in 1984 \cite{Riegert}.}
It should be pointed out that the structures $\Xi_n$ in \eqref{2.13} are independent for $n =1,2,3,4$. Specifically, it may be shown that
\begin{subequations}
\begin{align}
	\frac{1}{\sqrt{\nu \bar{\nu}}} \Xi_n = \frac{n}{\sqrt{\nu \bar{\nu}}} \Xi_1 
	- n(n-1)\Big( \mathfrak{A} - 2(n-2) \mathfrak{B} - (n-2)(n-3) \mathfrak{C} \Big)~,
\end{align}
where we have introduced the following conformally primary structures:
\begin{align}
	\mathfrak{A} &= \nabla^2 \Psi \nabla^2 \bar{\Psi}
	+ 2 \nabla^a \nabla^b \Psi \nabla_a \nabla_b \bar{\Psi} + \nabla^a \bar{\Psi} \Big(\nabla_a \nabla^2 \Psi + \tfrac 1 4 ( R_{ab} - \tfrac 1 6 \eta_{ab} R ) \nabla^b \Psi \Big)  \non \\
	&\phantom{=}~ - 2 \bar{\Psi} (\nabla^a \bar{\Psi} \nabla_a \Psi \nabla^2 \Psi + 2 \nabla^a \Psi \nabla^b \Psi \nabla_a \nabla_b \Psi) ~, \\
	\mathfrak{B} &= \Psi \Big( \nabla^2 \Psi \nabla^a \bar{\Psi} \nabla_a \bar{\Psi} + 2 \nabla^a \nabla^b \Psi \nabla_a \bar{\Psi} \nabla_b \bar{\Psi} \Big) ~, \\
	\mathfrak{C} &= \nabla^a \Psi \nabla_a \Psi \nabla^b \bar{\Psi} \nabla_b \bar{\Psi} ~.
\end{align}
\end{subequations}
Each of these structures may be written in a different form by making use of $\J \bar  \J =1$, in particular 
 $\mathfrak{C} = (\nabla^a  \Psi \nabla_a \bar \Psi )^2$.

So far we have considered only those primary deformation structures which do not 
contain the primary vector fields 
\bea 
\c_{\a\ad} := \nabla^\b{}_\ad V_{\a\b}~, \qquad \bar \c_{\a\ad} := \nabla_\a{}^\bd \bar V_{\ad \bd}~.
\label{218}
\eea
When considering deformations of Maxwell's theory, the structures containing $\c_a$ and $\bar \c_a$ would lead to contributions involving the classical equations of motion. 
This is the reason for avoiding such structures in the above discussion. 
However, making use of the vector fields \eqref{218} allows us to generate new deformations, including the following: 
\bea
\frac{ (\c \cdot \bar \c)^2} {\n \bar \n} ~, \qquad \frac{ (\c \cdot \c) (\bar \c \cdot \bar \c)^2} {\n \bar \n} ~, \qquad 
\frac{ (\c \cdot  \c)^2} {\n^2} ~, \qquad \frac{ (\bar \c \cdot \bar \c)^2} { \bar \n ^2} ~,
\eea
which may originate at the one-loop level.


\section{Elimination of auxiliary variables}
\label{Section3}

Given a duality-invariant theory described by the first-order action \eqref{AuxAction}, in which the self-coupling $ \mathfrak{S}^{\rm int}[ V] =  \mathfrak{S}^{\rm int}[ \n, \bar \n]$ obeys the condition \eqref{invariance2}, the final goal is to derive its reformulation $S[F]$ 
which is 
obtained by imposing the equation of motion \eqref{AuxEoM}. This equation is equivalent to 
\begin{align}
	V_{\a \b} = F_{\a \b} - \hf \frac{\d}{\d V^{\a \b}} 
		 \mathfrak{S}^{\rm int} [V]
	 \label{AuxEoM2}
\end{align}
and its conjugate. The latter equations can be solved in perturbation theory, say, within the loop expansion. 

Let us consider two examples. We start with a one-loop deformation in \eqref{LEEA}, and for simplicity we set 
$\hbar =1$, 
\begin{align}
 \mathfrak{L}^{\rm int}_\text{MM,def} = \k \sqrt{\n \bar \n} 
+ g \frac{ \big[ \square_c (\n\bar \n)^{ 1/8} \big]^2 } {  (\n\bar \n)^{ 1/4}}~. \label{2.9a} 
\end{align}
The equation of motion \eqref{AuxEoM2} takes the form
\begin{align}
		 V_{\a \b} = F_{\a \b} - V_{\a \b} \bigg \{ \frac{\k}{2} \Big ( \frac{\bar{\n}}{\n} \Big )^{\frac 1 2} 
		+ \frac{g \bar{\nu}}{4} \bigg [ (\nu \bar{\nu})^{-\frac 7 8} \square_c \bigg ( \frac{\square_c(\n \bar{\n})^{\frac 1 8}}{(\n \bar{\n})^{\frac 1 4}}\bigg )
		- (\nu \bar{\nu})^{- \frac 5 4} (\square_c (\nu \bar{\nu})^{\frac 1 8} )^2
		\bigg ]
		\bigg \} ~.
\end{align}
Eliminating the auxiliary fields gives 
\begin{align}
		L &= L_\text{MM} + g \O^{-\frac12} \big(\square_c \O^{\frac14}\big)^2 
		+ \frac{g^2\O^{- \frac 32}}{4  (1-(\k/2)^2)(1+(\k/2)^2)^2} \Big ( \Box_c (\O^{-\hf} \Box_c \O^{\frac 1 4}) - \O^{-\frac 3 4} (\Box_c \O^{\frac 1 4})^2 \Big )^2  \non \\
		&\phantom{=} \qquad \times  \bigg \{\Big(3-12(\k/2)^2 +20(\k/2)^4\Big) (\o +\bar{\o}) - 4(\k/2)\Big(2+\k/2-5(\k/2)^2 +2(\k/2)^3 \non \\
		& \qquad \qquad \qquad \quad  +9(\k/2)^4 +(\k/2)^5 \Big) \O  \bigg \}+\mathcal{O}(g^3)~,
\label{3.4}
\end{align}
where we have defined
\begin{align}
	\O = \frac{\big( 1 + (\k/2)^2 \big) (\o \bar{\o})^{\frac 1 2} - (\k/2) (\o + \bar{\o})}{\big( 1 - (\k/2)^2\big)^2} = 
		\hf (\cosh \g + 1)
	 \frac{\partial L_\text{MM}}{\partial \g}~.
\label{Omega}
\end{align}

Our second example is defined by 
\begin{align}
 \mathfrak{L}^{\rm int}_\text{MM,deformed} = \k \sqrt{\n \bar \n} 
+ g (\n\bar \n)^{ 1/8} \square_c (\n\bar \n)^{ 1/8}~. \label{2.9b}
\end{align}
The corresponding equation of motion for the auxiliary variable $V_{\a\b}$ is 
	\begin{align}
		 V_{\a \b} = F_{\a \b} - V_{\a \b} 
		\bigg \{ \frac{\k}{2} \Big ( \frac{\bar{\n}}{\n} \Big )^{\frac 1 2} 
		+ \frac{ g \bar{\nu}}{4} (\nu \bar{\nu})^{- \frac 7 8} \square_c (\nu \bar{\nu})^{\frac 1 8}
		\bigg \}~.
	\end{align}
Eliminating the auxiliary fields gives
\begin{align}
L &= L_\text{MM} +  g \O^{\frac14} \square_c \O^{\frac14} 
+ \frac{g^2 \O^{- \frac 3 2}  }{ 4  (1-(\k/2)^2)(1+(\k/2)^2)^2} \Big ( \Box_c \O^{\frac 1 4} \Big )^2  \non \\
&\phantom{=} \qquad \times \bigg \{ \Big(3-12(\k/2)^2 +20(\k/2)^4\Big) (\o +\bar{\o}) - 4(\k/2)\Big(2+\k/2-5(\k/2)^2 +2(\k/2)^3 \non \\
& \qquad \qquad \qquad \quad  +9(\k/2)^4 +(\k/2)^5 \Big) \O  \bigg \}+\mathcal{O}(g^3)~.
\label{3.8}
\end{align}

Both models \eqref{3.4} and \eqref{3.8} involve one and the same composite field $\O$, 
eq. \eqref{Omega}. What is the significance of $\O$? Let $\mathfrak{S} [F,V; g]$ be the action corresponding to the self-coupling \eqref{2.9a}, and $S[F;g]$ the $\sU(1)$ duality-invariant model which is obtained upon elimination of the auxiliary field 
$V_{ab}$. Since the parameter $g$ is inert under the $\sU(1)$ duality transformations, 
the functional 
\bea
\label{3.9}
\U(g) := \frac{\pa}{\pa g} S[F;g]
\eea
is duality invariant for any value of $g$ \cite{GZ2,GZ3,KT2}.\footnote{\label{FN}This property implies that, in perturbation theory, the leading contribution to the deformation of any self-dual theory must be invariant under the duality transformations of the original theory. We emphasise that this does not extend beyond first order; the sectors of \eqref{3.4} and \eqref{3.8} quadratic in $\hbar$ are not duality invariant.} In particular, $\U(g=0)$ is 
a duality-invariant functional in the ModMax theory.  As demonstrated in 
\cite{Ferko:2023wyi}, any two duality-invariant local observables $H_1 (F; \g)$ and 
$H_2(F; \g) $ are functionally dependent, and 
$ {\partial L_\text{MM}}/{\partial \g}$ is such an observable.\footnote{The energy-momentum tensor $T_{ab}$ in every $\sU(1)$ duality-invariant theory is duality invariant \cite{GR1,GZ2,GZ3}, and therefore every scalar duality-invariant observable $H(F)$ may be expressed as a function of $T_{ab}$. For the Mod-Max theory it holds that 
$ {\partial L_\text{MM}}/{\partial \g} = \hf \sqrt{T^{ab}T_{ab}}$, 
see \cite{Babaei-Aghbolagh:2022uij, FST-M1,FST-M2, Ferko:2023wyi} for the technical details and earlier references.}  
Thus it is natural to expect that the duality-invariant functional $\U(0)$ should be constructed in terms of $\O$. Analogous considerations apply to the $\sU(1)$ duality-invariant model generated by \eqref{2.9b}. 

In accordance with \cite{Ferko:2023wyi}, every duality-invariant scalar observable $\cO(F)$ in the ModMax theory can be expressed as a function of $\O$. However, this is no longer the case if we allow for functionals involving derivatives of the field strength $F_{ab}$.
Let us consider an infinitesimal duality transformation in the ModMax theory 
\begin{align} 
\d_\vf F_{\a\b } = \ri \vf F_{\a\b} \left( \cosh \g -  \sqrt{ \frac{\bar \o}{\o} }\sinh \g \right)
~ \implies ~ \d_\vf \o = 2\ri \vf \Big( \o \cosh \g - \sqrt{\o\bar \o} \sinh \g \Big)~.
\end{align} 
Introducing 
\bea
I :=  \sqrt{\o} (1+\cosh \g) - \sqrt{\bar \o} \sinh \g ~\implies ~ I \bar I = 4\O~, 
\label{3.11}
\eea
we then observe that\footnote{This transformation law follows from the auxiliary variable formulation. Specifically, the equation of motion for $V_{ab}$ in ModMax theory implies that $(1+\cosh \g)\sqrt{\n} = I$. Since $\d_\vf \nu = 2 \ri \nu$ under $\sU(1)$ duality transformations, we obtain equation \eqref{3.12}.}
\bea
\label{3.12}
\d_\vf I = \ri \vf I~.
\eea
This result immediately implies that $\O$ is duality invariant. Moreover, it also implies the existence of new primary and duality-invariant observables, such as $I (\square_c \sqrt{\bar I})^2$, which are functionally independent of $\O$.

\section{Higher-derivative deformations of super ModMax}
\label{Section4}

It is also of interest to study consistent higher-derivative deformations of 
the $\cN=1$ supersymmetric extension of the ModMax theory which was discovered in  \cite{Bandos:2021rqy} and independently re-derived in \cite{K21} 
\bea
S_{\rm SMM}[W,{\bar W}] &=&
\frac{1}{4} \cosh \g \int  \rd^4 x \rd^2 \q  \,\cE \, W^2 +{\rm c.c.}
+ \frac{1}{4}\sinh \g   \int \rd^4 x \rd^2 \q \rd^2\bar \q \,E \,
\frac{W^2\,{\bar W}^2}{\sqrt{u\bar u} }~,
\label{superMM}
\eea
where $\cE$ denotes the chiral integration measure.
This theory is a unique representative in the family of $\sU(1)$ duality-invariant nonlinear theories for the $\cN=1$ vector multiplet of the general form \cite{KT1,KT2,KMcC,KMcC2}
\bea
S[W,{\bar W};\U] &=&
\frac{1}{4} \int  \rd^4 x \rd^2 \q  \,\cE \, W^2 +{\rm c.c.}
+ \frac14  \int \rd^4 x \rd^2 \q \rd^2\bar \q \,E \,
\frac{W^2\,{\bar W}^2}{\U^2}\,
\L\left(\frac{u}{\U^2},
\frac{\bar u}{\U^2}\right)~,
\label{superED}
\eea
where $W^2 =W^\a W_\a$ and $\bar W^2 = \bar W_\ad \bar W^\ad$,  the variable $u$ is defined by
\bea
u  := \frac{1}{8} (\cD^2 - 4  \bar R)  W^2~,
\eea
the compensator $\U$ is a nowhere vanishing real scalar with the super-Weyl transformation 
\bea
\d_\s \U = (\s +\bar \s) \U~, \qquad \bar \cD_\bd \s=0~, 
\label{compensator}
\eea
and $\L(\o, \bar \o) $ obeys the self-duality equation \eqref{GZ4}.\footnote{The analyses in Refs. \cite{KT1,KT2,KMcC,KMcC2} were restricted to the case that the self-coupling $\L(\o,\bar \o)$ must be a real analytic function in a neighbourhood of $\o=0$.} 
We recall that the infinitesimal super-Weyl transformation of the spinor covariant derivatives \cite{HT} is given by 
\label{superweyl}
\bea
\d_\s \cD_\a &=& ( {\bar \s} - \hf \s)  \cD_\a + \cD^\b \s \, M_{\a \b}  ~, \qquad
\d_\s \bar \cD_\ad  =  (  \s -  \hf {\bar \s})
\bar \cD_\ad +   \bar \cD^\bd  {\bar \s}   {\bar M}_{\ad \bd} ~,
\eea
see \cite{Kuzenko:2022skv} for a review.
The transformation law \eqref{compensator} implies that \eqref{superED} is super-Weyl invariant.
We also remind  the reader that the super-Weyl
transformation law of the chiral field strength $W_\a$ is 
\bea
\d_\s W_\a = \frac{3}{2} \s \, W_\a \quad \implies 
\quad \d_\s (\cD^\a W_\a ) = (\s +\bar \s) \cD^\a W_\a~,
\eea
and therefore the following composite antichiral scalar 
\bea
\bm{u}
:= \frac 18 (\cD^2 - 4 {\bar R}) \Big( \frac{W^2 }{ \U^2 } \Big)
\label{A.7}
\eea
is super-Weyl invariant.  What singles out the super ModMax theory \eqref{superMM}, is that the corresponding $\L(\o,\bar \o)$ is given by \eqref{LambdaMM} and thus the $\U$-dependence drops out.  As a consequence, the action is locally superconformal.\footnote{The super ModMax theory has been studied within the framework of $T\bar T$-like deformations, see \cite{FST-M2} and references therein. Higher-spin extensions of the ModMax theory were given in \cite{Kuzenko:2021qcx}. The $\sSL(2,{\mathbb R})$ duality-invariant coupling of this theory to the dilaton-axion multiplet was described in \cite{K21} following the general formalism of \cite{KT2, KMcC}.
}

Given a $\sU(1)$ duality-invariant vector multiplet model with action $S[W,{\bar W};\U]$, there exists its
reformulation with auxiliary variables \cite{K13,ILZ}
\begin{align} 
	\mathfrak{S}[W,\bar W, \eta, \bar \eta; \U]&= \int{\rm d}^4 x \rd^2\q\,\cE \, \Big\{ \eta W -\hf \eta^2 - \frac{1}{4} W^2\Big\} 
	+{\rm c.c.} + \mathfrak{S}^{\rm int} [ \eta, \bar \eta; \U] ~,
	\label{superMM-aux}
\end{align}
where the auxiliary spinor $\eta_\a$ 
is only constrained to be covariantly chiral, $\bar \cD_\bd \eta_\a =0$.
We assume that $\eta_\a$ and its conjugate $\bar \eta_\ad$ are auxiliary  
superfields in the sense that the equation of motion for $\eta_\a$,
\bea
W_\a= \eta_\a- 2\frac{\d}{\d \eta^\a}  {\mathfrak S}^{\rm int} [\eta, \bar \eta;\U]~,
\eea
and its conjugate may be solved to express $\eta_\a$ as a functional 
of the field strength $W_\a$ and its conjugate, $\eta_\a =\eta_\a [W, \bar W;\U]$.
As a result, we end up with the action 
\bea
S[W,\bar W;\U] = \mathfrak{S}[W,\bar W, \eta, \bar \eta;\U]\Big|_{\eta_\a =\eta_\a [W, \bar W;\U]}~,
\eea
which describes the dynamics of the vector multiplet. 
The action \eqref{superMM-aux} defines a $\sU(1)$ duality-invariant system provided  
${\mathfrak S}^{\rm int} [\eta, \bar \eta;\U]$ is invariant under rigid $\sU(1)$  transformations,
\bea
\eta_\a \to   \re^{\ri \vf}  \eta_\a, \qquad 
\bar \eta_\ad \to \re^{-\ri \vf} \bar \eta_\ad ~, 
\qquad \vf \in {\mathbb R}~.
\label{RU(1)}
\eea
This implies that the action $S[W,\bar W;\U] $ obeys the self-duality equation \cite{KT1,KT2,KMcC}
\begin{subequations}\label{SDE-super}
\bea
{\rm Im} \int \rd^4 x \rd^2 \q  \,\cE \Big\{ W^\a W_\a  +M^\a M_\a \Big\} =0~,
\eea
where we have introduced 
\bea
{\rm i}\,M_\a := 2\, \frac{\d }{\d W^\a}\,S[W , {\bar W}]~,  \qquad \bar \cD_\bd M_\a=0~.
\eea
\end{subequations}
In the relations \eqref{SDE-super}, $W_\a$ is chosen to be a general chiral spinor superfield.

It is worth pointing out that the self-coupling $\mathfrak{S}^{\rm int} [ \eta, \bar \eta; \U] $ corresponding to the duality-invariant nonlinear supersymmetric electrodynamics 
\eqref{superED} is 
\begin{align}
	\mathfrak{S}^{\rm int} [ \eta, \bar \eta; \U] 
	&= \frac{1}{4} \int \rd^4 x \rd^2 \q \rd^2\bar \q \,E \, \frac{\eta^2 \bar \eta^2}{\U^2} 
	{\mathfrak F}\Big( \frac{v}{\U^2} , \frac{\bar v}{\U^2} \Big)~,
	\qquad
	v:= \frac{1}{8} (\cD^2 -4\bar R) \eta^2~,
	\label{SMM4.14}
\end{align}
This action is super-Weyl invariant provided $\eta_\a$ transforms as
\bea
\d_\s \eta_\a = \frac{3}{2} \s \eta_\a \quad \implies 
\quad \d_\s (\cD^\a \eta_\a ) = (\s +\bar \s) \cD^\a \eta_\a~.
\eea
The model is $\sU(1)$ duality invariant if
${\mathfrak F} ( v,  \bar v) ={\mathfrak F} (v \bar v)$, 
see \cite{K13} for the technical details. In the super ModMax case, it holds   \cite{K21}
that 
\bea
\mathfrak{S}^{\rm int}_{\rm SMM} [ \eta, \bar \eta] 
= \frac{\k}{4} \int \rd^4 x \rd^2 \q \rd^2\bar \q \,E \, \frac{\eta^2 \bar \eta^2}{\sqrt{v\bar v}} ~,
\label{SMM-int}
\eea
where $\k$ is related to the ModMax coupling constant $\g$ as in \eqref{MM2.4}.
In accordance with \eqref{SMM4.14}, this functional has no dependence on
the compensator $\U$. 

Let us consider the in-out vacuum amplitude for the super ModMax theory 
\bea
Z = \int [\mathfrak{D} V] [\mathfrak{D} \eta_\a ] [\mathfrak{D} \bar \eta_\ad ] \,\d_+\big[ \k(V) \big ] \d_- \big[\bar \k (V) \big]{\rm Det}(\cH ) \,
\exp \left\{ \frac{\ri} {\hbar} \mathfrak{S}_{\rm SMM} [W,\bar W, \eta, \bar \eta] \right\}~,
\label{in-out-super}
\eea
where $V=\bar V$ is the gauge superfield which determines $W_\a$ as the following descendant
\bea
W_\a = -\frac 14 (\bar \cD^2 - 4R) \cD_\a V~,
\eea
and $\k(V)$ is the gauge fixing function
\bea
\k(V) = - \frac 14 (\bar \cD^2 - 4R)V + \x ~, \qquad \bar \cD_\ad \x = 0~,
\eea
with $\x$ being a background chiral superfield. Finally, $\cH$ denotes the Faddeev-Popov operator
\begin{align}
	\cH = \left(
	\begin{array}{cc}
		0 & -\frac14 (\bar \cD^2 -4R) \\
		-\frac14 (\cD^2 - 4 \bar R) & 0 \\
	\end{array}
	\right)~,
\end{align}
which acts on the space of chiral ($\f$) and antichiral ($\bar \f$) pairs. More details about the covariant quantisation of the vector multiplet in a supergravity background can be found, e.g., in 
\cite{Buchbinder:1998qv}.

In accordance with \eqref{superMM-aux} and \eqref{SMM-int}, 
the functional $\hbar^{-1} \mathfrak{S}_{\rm SMM}[W, \bar W, \eta, \bar \eta]$ is invariant under rigid re-scalings 
\bea
\hbar \to \l^2 \hbar ~, \qquad W_\a(z) \to \l W_\a (z)~, \qquad \eta_{\a}(z) \to \l \eta_{\a}(z)~.
\label{scale2}
\eea
Modulo quantum anomalies, the effective action is expected to respect this scale symmetry.  
We are interested in a local deformation of the super ModMax theory
\bea
{\G}_{\rm SMM} [W, \bar W, \eta, \bar \eta] &=& 
\mathfrak{S}_{\rm SMM} [W, \bar W, \eta, \bar \eta]  
+ \sum_{n=1}^{\infty} \hbar^{n} \G^{(n)} [ \eta, \bar \eta] 
\non \\
& \equiv & 
\mathfrak{S}_{\rm SMM} [W,\bar W,\eta, \bar \eta]  +
\D \G_{\rm SMM} [ \eta, \bar \eta] 
\eea
with the following properties: (i) the functional ${\hbar}^{-1} \D \G_{\rm SMM} [ \eta, \bar \eta] $
is invariant under \eqref{scale2}; (ii) $\D \G_{\rm SMM} [ \eta, \bar \eta] $ is super-Weyl invariant; and 
(iii) $\D \G_{\rm SMM} [ \eta, \bar \eta] $ is invariant under the $\sU(1)$ transformations \eqref{RU(1)}.
These properties imply that the deformation $\D \G_{\rm SMM} [ \eta, \bar \eta] $ does not contain the  self-coupling \eqref{SMM-int} corresponding to the super ModMax theory. 

To understand the structure of allowed contributions to  $\D \G_{\rm SMM} [ \eta, \bar \eta] $, techniques are required to generate primary descendants of $\eta_\a$ and $\bar \eta_\ad$. Such methods are available only in the presence of a compensator, for instance $ (\cD^2 - 4 {\bar R}) \big( {\eta^2 }\U^{-2}  \big)$ is a primary dimension-zero antichiral scalar. However, since we are interested in locally superconformal deformations, 
no supergravity compensator can be present in the deformed action. The only remaining option is to allow for $\U$ being a descendant of 
the vector multiplet, which means that\footnote{This idea has been used in \cite{Kuzenko:2019vaw, Kuzenko:2023igt}.} 
\bea
\U = \cD \eta \bar \cD \bar \eta ~.
\eea
However, in the ModMax theory the $\q$-independent component of $\cD \eta$ vanishes on-shell  \cite{Bandos:2021rqy}. Thus we are allowed to take into account only those structures which involve $\U$ in the numerator.   

Let us describe two families of functionals that can be used to generate consistent deformations of the super ModMax theory.  One of them involves
the supersymmetric Fradkin-Tseytlin operator \cite{ButterK13,BdeWKL}, $\D$ which is defined by
\bea
\D\bar \F := -\frac{1}{64} (\bar \cD^2 -4R ) \Big\{ \cD^2 \bar \cD^2 \bar \F 
+ 8 \cD^\a (G_{\a\ad}\bar \cD^\ad \bar \F)\Big\}~, \qquad 
\bar \cD_\ad \D\bar \F =0
\label{Delta}
\eea
and has the super-Weyl transformation law
\bea
\d_\s \D\bar \F = 3\s \D\bar \F ~,
\label{sW-Delta}
\eea
provided $\bar \F$ is inert under the super-Weyl transformations, $\d_\s \bar \F =0$.

Using $\eta_\a$ and $\bar \eta_\ad$ allows us to construct a primary dimension-zero antichiral scalar 
\bea
\bm{v}:= \frac 18 (\cD^2 - 4 {\bar R}) \Big( \frac{\eta^2 }{ \cD \eta \bar \cD \bar \eta } \Big)
\eea
and its conjugate $\bar {\bm v}$. Now we can generate primary dimension-three chiral scalars of the form 
\bea
\bar{\bm v}^{-n} \D  {\bm v}^{-n}~, \qquad n>0~,
\eea
each of which can be integrated over the chiral subspace to result in a super-Weyl invariant  
\bea
\int{\rm d}^4 x \rd^2\q\,\cE \, \bar {\bm v}^{-n} \D  {\bm v}^{-n}~.
\label{4.26}
\eea
This functional is obviously invariant under rigid $\sU(1)$ transformations
\eqref{RU(1)}.
Another type of consistent deformation is realised as an integral over the full superspace
\bea
- \frac 18 \int \rd^4 x \rd^2 \q \rd^2\bar \q \,E \, H_1 ({\bm v} \bar {\bm v})
\cD^\a (\bar \cD^2 - 4R) \cD_\a H_2 ({\bm v} \bar {\bm v})~.
\label{4.27}
\eea

Above we have described two families of higher-derivative deformations of the super ModMax theory enjoying both $\sU(1)$ duality invariance and super-Weyl symmetry. These are determined by the functionals \eqref{4.26} and \eqref{4.27}. They are of particular interest since they are also symmetric under the re-scaling transformation \eqref{scale2}, hence they may appear in the low-energy effective action for super ModMax theory. It should be noted, however, that these deformations do not describe supersymmetric extensions of those presented in section \ref{Section2} due to their dependence on $\cD \eta$.

It is possible to construct a supersymmetric generalisation of the primary field $\J$ introduced in \eqref{2.13}.
It makes use of the primary dimensionless scalar 
\begin{align}
	\bm{w} := \frac{\bm{v}}{\bar{\bm{v}}}~,
\end{align}
which may admit a well-defined $\cD \eta\to 0$ limit in its bosonic sector. To probe this further, we take the flat-space limit, $\cD_A \rightarrow D_A$, and find that $\bm{w}$ takes the form
\begin{align}
	\bm{w} &= \frac{D^2 \eta^2}{\bar{D}^2 \bar{\eta}^2} \bigg(1 - 2 \frac{D^\a \eta^2 D_\a (D \eta \bar{D} \bar{\eta})}{D^2 \eta^2 D\eta \bar{D} \bar{\eta}} + 2 \frac{\eta^2 D^2 (D \eta \bar{D} \bar{\eta})}{D^2\eta^2 (D\eta \bar{D} \bar{\eta})^2} \bigg) \bigg( 1 + 2 \frac{\bar{\eta}^2 \bar{D}^2(D\eta \bar{D} \bar{\eta})}{\bar{D}^2\bar{\eta}^2 (D\eta \bar{D} \bar{\eta})^2} \bigg)^{-1} \non \\
	 &\qquad \qquad \times \bigg ( 1 - 2 \bar{D}_\ad \bar{\eta}^2 \bar{D}^\ad (D \eta \bar{D} \bar{\eta}) \bigg[ 1 + 2 \frac{\bar{\eta}^2 \bar{D}^2 (D \eta \bar{D} \bar{\eta})}{\bar{D}^2 \bar{\eta}^2 (D \eta \bar{D} \bar{\eta})^2} \bigg]^{-1} \non \\
	 & \qquad \qquad \qquad \qquad - 2 \big(\bar{D}_\ad \bar{\eta}^2 \bar{D}^\ad (D \eta \bar{D} \bar{\eta})\big)^2 \bigg[ 1 + 2 \frac{\bar{\eta}^2 \bar{D}^2 (D \eta \bar{D} \bar{\eta})}{\bar{D}^2 \bar{\eta}^2 (D \eta \bar{D} \bar{\eta})^2} \bigg]^{-2}\bigg )~.
\end{align}
Taking the $\q$-independent component of $\bm{w}$ and setting all fermionic fields to zero, the resulting expression is simply
\begin{subequations}
\begin{align}
	 \label{4.31a}
	 \bm{w}|_{\q = 0} = \frac{\n -\hf  \t^2}{\bar{\n}- \hf \bar{\t}^2} ~,  
\end{align}
where we have made the definitions\footnote{These definitions agree with eq. (4.23) in \cite{KT2} in the case that $\eta_\a = W_\a$.}
\begin{align}
	V_{\a \b} = -\frac{\ri}{2} D_{(\a} \eta_{\b)}|_{\q = 0} ~, \qquad \t = -\hf D^{\a} \eta_{\a} |_{\q = 0}~.
\end{align}
\end{subequations}
Finally, taking the legal limit $\t \rightarrow 0$ in eq. \eqref{4.31a}, we find that $\bm{w}|_{\q = 0} = \Psi$, see eq. \eqref{2.13}.
This analysis indicates that a functional of the form
\bea
\label{4.32}
- \frac 18 \int \rd^4 x \rd^2 \q \rd^2\bar \q \,E \, \bar{\bm{w}}^n
\cD^\a (\bar \cD^2 - 4R) \cD_\a \bm{w}^n~,
\eea
defines a superconformal and duality-invariant deformation of the super ModMax action. 
However, a separate analysis is required to investigate whether this action is non-singular in the bosonic sector, since so far we have only analysed the component structure of $\bm w$.


\section{Concluding comments} 
\label{Section5}
In this paper we have described families of higher-derivative deformations of the ModMax theory \eqref{2.7}. The general form of such an action is 
\bea
\mathfrak{S} [F,V] = \mathfrak{S}_{\rm MM} [F,V] + 
\mathfrak{H} (\S, \U, \bar \U, \Xi_n, \bar \X_n)~,
\eea
where the primary composite fields $\S$, $\U$ and $\X_n$ are defined in the relations \eqref{2.12} and \eqref{2.13}.
While there is quite a large freedom in the choice of such extensions, significant restrictions occur if one requires that they are invariant under the rescalings \eqref{scale}. The presence of this symmetry is key to our analysis as we expect it to be a feature of the quantum effective action \eqref{2.11a}. To this end, we now utilise the techniques developed above to construct an ansatz for the one-loop deformation of the ModMax theory. 

We choose the self-coupling in the first-order action \eqref{AuxAction} to be
\begin{subequations}
\begin{align}
 \mathfrak{S}^{\rm int} [\n,\bar{\n}] &= \hbar \int \rd^4 x \,e \, \sqrt{\nu \bar{\nu}}
 								\Big \{ 
 								g_1 \U^2 + \bar{g}_1 \bar{\U}^2  
 								+ g_2 \U \bar{\U} 
 								+ \sum_{n=1}^{4} g_3^{(n)} \Xi_n + g_4 \S^2 
 								\Big \}
 								\\
 							 &= \hbar \int \rd^4 x \,e \,
 								\Big \{ 
 								\frac{g_1 \bar{\n}^{\frac 1 2} (\Box_c \n^{\frac 1 4})^2  
 									+ \bar{g}_1 {\n}^{\frac 1 2} (\Box_c \bar{\n}^{\frac 1 4})^2}{(\n \bar{\n})^{\frac 1 2}}
 								+ g_2 \frac{\Box_c \n^{\frac 1 4} \Box_c \bar{\n}^{\frac 1 4}}{(\n \bar{\n})^{\frac 1 4}}
 								\non \\
 							 & \qquad \qquad \qquad
 								+ \sum_{n=1}^{4} g_3^{(n)} \bar{\Psi}^n \D_0 \Psi^n + g_4 \frac{(\Box_c(\n \bar{\n})^{\frac 1 8})^2}{(\n \bar{\n})^{\frac 1 4}}
 								\Big \}~,
\end{align}
\end{subequations}
where $g_1 \in \mathbb{C}$ and $g_2,g_3^{(n)},g_4 \in \mathbb{R}$.
Eliminating the auxiliary fields to leading order in $\hbar$ via the equation of motion \eqref{AuxEoM2} leads to the higher-derivative Lagrangian
\begin{align}
	L = L_\text{MM} &+ \hbar \bigg \{ \frac{g_1 \bar{I} (\Box_c \sqrt{I})^2 + \bar{g}_1 {I} (\Box_c \sqrt{\bar{I}})^2 }{2 \O} + g_2 \frac{\Box_c \sqrt{I} \Box_c \sqrt{\bar{I}}}{\sqrt{2\O}} + \sum_{n=1}^{4} g_3^{(n)} \frac{I^{2n}}{\bar{I}^{2n}} \D_0 \frac{\bar{I}^{2n}}{I^{2n}} \non \\
	& \qquad \qquad  + g_4 \O^{-\frac12} \big(\square_c \O^{\frac14}\big)^2 \bigg \} + \mathcal{O}(\hbar^2) ~.
\label{4.3}
\end{align}
In accordance with footnote \ref{FN}, the sector of $L$ linear in $\hbar$ must be duality invariant. This property immediately follows from 
the transformation law \eqref{3.12}  in conjunction with the duality invariance of $\O$,
eq. \eqref{3.11}. All structures in \eqref{4.3} may contribute to the logarithmically divergent part of the one-loop effective action for the ModMax theory. 
However, explicit calculation of one-loop divergences in the ModMax theory remain to be completed \cite{GKP}.

The main constructions of this paper can be extended to the conformal chiral two-form field theory in six dimensions \cite{Bandos:2020hgy} within the approach of \cite{Ferko:2024zth}, which 
combined the virtues of the six-dimensional Pasti-Sorokin-Tonin formulation \cite{Pasti:1996vs, Pasti:1997gx} with the four-dimensional formalism developed by Ivanov, Nurmagambetov and Zupnik \cite{Ivanov:2014nya}.
Higher-derivative deformations of the conformal chiral two-form field theory in six dimensions 
can also be studied using the approach by Mkrtchyan et al.
\cite{Mkrtchyan:2019opf, Bansal:2021bis, Avetisyan:2022zza}.

In section \ref{Section4} we have studied higher-derivative deformations of the super ModMax theory. Their structure is more restrictive than in the non-supersymmetric case. Moreover, they do not look ``natural'' due to built-in singularities.  In this sense explicit calculations of loop quantum corrections may be revealing.

Unlike the purely bosonic case, there exist consistent deformations of the super ModMax theory without higher derivatives that preserve both the $\sU(1)$ duality invariance and super-Weyl symmetry, such as the following:
\bea
\mathfrak{S}&=&\mathfrak{S}_{\rm SMM} [W, \bar W, \eta, \bar \eta]  
+ \frac{1}{4} \int \rd^4 x \rd^2 \q \rd^2\bar \q \,E \, \frac{\eta^2 \bar \eta^2}{\cD \eta \bar \cD \bar \eta}  \sum_{n=1}^{\infty}  g_n (\bar{\bm v} 
{\bm v})^{-n} \non \\
&=&	\mathfrak{S}_{\rm SMM} [W, \bar W, \eta, \bar \eta]  
+  \frac{1}{4}  \int \rd^4 x \rd^2 \q \rd^2\bar \q \,E \, {\eta^2 \bar \eta^2}
\sum_{n=1}^{\infty} g_n
\frac{(\cD \eta \bar \cD \bar \eta)^{2n-1}}{(\bar{v}  {v})^{n}}~.
\label{5.4}
\eea
In a flat superspace background, the purely bosonic sector of the deformation
in \eqref{5.4}  is 
\begin{align}
 \int \rd^4 x \, \sum_{n=1}^{\infty} g_n
   \frac{(4\t \bar{\t})^{2n-1}}{\big[ (\n - \hf \t^2)(\bar{\n} - \hf \bar{\t}^2) \big]^{n-1}}~.
\end{align}
Upon elimination of the auxiliary field $V_{ab}$ in the perturbation theory,  
the equation of motion for the auxiliary field $D$ of the vector multiplet has a solution
$D=0$, as in the ModMax theory \cite{BLST}.

\noindent
{\bf Acknowledgements:}  We are grateful to Dmitri Sorokin and Arkady Tseytlin for useful comments and suggestions. This work was supported in part by the Australian Research Council, project No. DP230101629.

\begin{footnotesize}

\end{footnotesize}


\begin{thebibliography}{66}

\bibitem{BLST}
I.~Bandos, K.~Lechner, D.~Sorokin and P.~K.~Townsend,
``A non-linear duality-invariant conformal extension of Maxwell's equations,''
Phys. Rev. D \textbf{102}, 121703 (2020)
[arXiv:2007.09092 [hep-th]].

\bibitem{B-B} I. Bialynicki-Birula, ``Nonlinear electrodynamics: Variations on a theme by Born and Infeld,'' in {\it Quantum Theory of Particles and Fields}, 
B. Jancewicz and J. Lukierski (Eds.), 
World Scientific, 1983, pp. 31--48. 

\bibitem{Kosyakov}
B.~P.~Kosyakov,
``Nonlinear electrodynamics with the maximum allowable symmetries,''
Phys. Lett. B \textbf{810}, 135840 (2020)
[arXiv:2007.13878 [hep-th]].


\bibitem{GZ1}
M. K.~Gaillard and B.~Zumino,
``Duality rotations for interacting fields,''
Nucl.\ Phys.\  {\bf B193},  221 (1981). 

\bibitem{GR1}
G.~W.~Gibbons and D.~A.~Rasheed,
``Electric-magnetic duality rotations in nonlinear electrodynamics,''
Nucl.\ Phys.\  {\bf B454}, 185 (1995) 
[arXiv:hep-th/9506035].

\bibitem{GR2}
G. W.~Gibbons and D. A.~Rasheed,
``SL(2,R) invariance of non-linear electrodynamics
coupled to an axion and a dilaton,''
Phys.\ Lett.\  {\bf B365}, 46 (1996) 
[hep-th/9509141].

\bibitem{GZ2}
M.~K.~Gaillard and B.~Zumino,
``Self-duality in nonlinear electromagnetism,''
in {\it Supersymmetry and Quantum Field Theory},
J.~Wess and V.~P.~Akulov (Eds.), Springer Verlag, 1998, pp. 121--129 [arXiv:hep-th/9705226].

\bibitem{GZ3}
M.~K.~Gaillard and B.~Zumino,
``Nonlinear electromagnetic self-duality
and Legendre transformations,'' in {\it Duality and
	Supersymmetric Theories}, D.~I.~Olive and
P.~C.~West (Eds.), Cambridge University Press,
1999, pp. 33--48 [hep-th/9712103].


\bibitem{K21}
S.~M.~Kuzenko,
``Superconformal duality-invariant models and $\mathcal{N} = 4$ SYM effective action,''
JHEP \textbf{09}, 180 (2021)
[arXiv:2106.07173 [hep-th]].


\bibitem{IZ_N3} 
  E.~A.~Ivanov and B.~M.~Zupnik,
  ``N=3 supersymmetric Born-Infeld theory,''
  Nucl.\ Phys.\ B {\bf 618}, 3 (2001)
  [hep-th/0110074].



\bibitem{IZ1} 
  E.~A.~Ivanov and B.~M.~Zupnik,
  ``New representation for Lagrangians of self-dual nonlinear electrodynamics,''
 in {\it Supersymmetries and Quantum Symmetries. Proceedings of the 16th Max Born Symposium, SQS'01: Karpacz, Poland, September 21--25, 2001}, E. Ivanov (Ed.), Dubna, 2002, pp. 235--250 
 [hep-th/0202203].

\bibitem{IZ2} 
  E.~A.~Ivanov and B.~M.~Zupnik,
  ``New approach to nonlinear electrodynamics: Dualities as symmetries of interaction,''
  Phys.\ Atom.\ Nucl.\  {\bf 67}, 2188 (2004)
  [Yad.\ Fiz.\  {\bf 67}, 2212 (2004)]
  [hep-th/0303192].

\bibitem{Hatsuda:1999ys}
M.~Hatsuda, K.~Kamimura and S.~Sekiya,
``Electric magnetic duality invariant Lagrangians,''
Nucl. Phys. B \textbf{561}, 341 (1999)
[arXiv:hep-th/9906103 [hep-th]].

\bibitem{Born:1934gh}
M.~Born and L.~Infeld,
``Foundations of the new field theory,''
Proc. Roy. Soc. Lond. A \textbf{144} no. 852, 425 (1934).

\bibitem{Bandos:2021rqy}
I.~Bandos, K.~Lechner, D.~Sorokin and P.~K.~Townsend,
``ModMax meets Susy,''
JHEP \textbf{10}, 031 (2021)
[arXiv:2106.07547 [hep-th]].

\bibitem{KT1}
S.~M.~Kuzenko and S.~Theisen,
``Supersymmetric duality rotations,''
JHEP {\bf 0003}, 034 (2000)
[arXiv:hep-th/0001068].

\bibitem{KT2}
S.~M.~Kuzenko and S.~Theisen,
``Nonlinear self-duality and supersymmetry,''
Fortsch.\ Phys.\  {\bf 49}, 273 (2001) [arXiv:hep-th/0007231].

\bibitem{Babaei-Aghbolagh:2022itg}
H.~Babaei-Aghbolagh, K.~Babaei Velni, D.~M.~Yekta and H.~Mohammadzadeh,
``Manifestly SL(2, R) duality-symmetric forms in ModMax theory,''
JHEP \textbf{12}, 147 (2022)
[arXiv:2210.13196 [hep-th]].

\bibitem{Banerjee:2022sza}
A.~Banerjee and A.~Mehra,
``Maximally symmetric nonlinear extension of electrodynamics with Galilean conformal symmetries,''
Phys. Rev. D \textbf{106}, no.8, 085005 (2022)
[arXiv:2206.11696 [hep-th]].

\bibitem{FT}
E.~Fradkin and A.~A.~Tseytlin,
``Quantum equivalence of dual field theories,''
Annals Phys. \textbf{162}, 31 (1985).

\bibitem{RT}
R.~Roiban and A.~Tseytlin,
``On duality symmetry in perturbative quantum theory,''
JHEP \textbf{10}, 099 (2012)
[arXiv:1205.0176 [hep-th]].

\bibitem{Pinelli} J. R. Pinelli, ``Heat kernel techniques for ModMax electrodynamics: A conformal duality-invariant extension of Maxwell electrodynamics,'' Master's Thesis, The University of Western Australia, 2021. 

\bibitem{Ebert:2024zwv}
S.~Ebert, C.~Ferko, C.~L.~Martin and G.~Tartaglino-Mazzucchelli,
``Flows in the space of interacting chiral boson theories,''
[arXiv:2403.18242 [hep-th]].


\bibitem{AFZ}
P.~Aschieri, S.~Ferrara and B.~Zumino,
``Duality rotations in nonlinear electrodynamics and in extended supergravity,''
  Riv.\ Nuovo Cim.\  {\bf 31}, 625 (2008)
  [arXiv:0807.4039 [hep-th]].


\bibitem{Chemissany:2011yv} 
  W.~Chemissany, R.~Kallosh and T.~Ortin,
  ``Born-Infeld with higher derivatives,''
  Phys.\ Rev.\ D {\bf 85}, 046002 (2012)
  [arXiv:1112.0332 [hep-th]].

\bibitem{AF} 
  P.~Aschieri and S.~Ferrara,
  ``Constitutive relations and Schroedinger's formulation of nonlinear electromagnetic theories,''
  JHEP {\bf 1305}, 087 (2013)
  [arXiv:1302.4737 [hep-th]].

\bibitem{AFT} 
  P.~Aschieri, S.~Ferrara and S.~Theisen,
  ``Constitutive relations, off shell duality rotations and the hypergeometric form of Born-Infeld theory,''
  Springer Proc.\ Phys.\  {\bf 153}, 23 (2014)
  [arXiv:1310.2803 [hep-th]].

\bibitem{Carrasco:2011jv}
J.~J.~M.~Carrasco, R.~Kallosh and R.~Roiban,
``Covariant procedures for perturbative non-linear deformation of duality-invariant theories,''
Phys. Rev. D \textbf{85}, 025007 (2012)
[arXiv:1108.4390 [hep-th]].

\bibitem{Broedel:2012gf}
J.~Broedel, J.~J.~M.~Carrasco, S.~Ferrara, R.~Kallosh and R.~Roiban,
``N=2 Supersymmetry and U(1)-Duality,''
Phys. Rev. D \textbf{85}, 125036 (2012)
[arXiv:1202.0014 [hep-th]].

\bibitem{Ivanov:2012bq}
E.~A.~Ivanov and B.~M.~Zupnik,
``Bispinor auxiliary fields in duality-invariant electrodynamics revisited,''
Phys. Rev. D \textbf{87}, no.6, 065023 (2013)
[arXiv:1212.6637 [hep-th]].

\bibitem{ILZ2}
E.~Ivanov, O.~Lechtenfeld and B.~Zupnik,
``New approach to duality-invariant nonlinear electrodynamics,''
J. Phys. Conf. Ser. \textbf{474}, 012023 (2013)
[arXiv:1310.5362 [hep-th]].


\bibitem{BN} 
  G.~Bossard and H.~Nicolai,
 ``Counterterms vs. dualities,''
  JHEP {\bf 1108}, 074 (2011)
  [arXiv:1105.1273 [hep-th]].

\bibitem{WB} J.~Wess and J.~Bagger,
{\it Supersymmetry and Supergravity},
Princeton Univ. Press, 1992.


\bibitem{Buchbinder:1998qv} I.~L.~Buchbinder and S.~M.~Kuzenko,
{\it Ideas and Methods of Supersymmetry and
Supergravity or a Walk Through Superspace}
Bristol, UK: IOP (1998) 656 p.

\bibitem{Kuzenko:2022skv}
S.~M.~Kuzenko, E.~S.~N.~Raptakis and G.~Tartaglino-Mazzucchelli,
``Superspace Approaches to $\mathscr {N} = \text{1}$ Supergravity,''
doi:10.1007/978-981-19-3079-9\_40-1
[arXiv:2210.17088 [hep-th]].


\bibitem{Kuzenko:2019nlm}
S.~M.~Kuzenko,
``Manifestly duality-invariant interactions in diverse dimensions,''
Phys. Lett. B \textbf{798}, 134995 (2019)
[arXiv:1908.04120 [hep-th]].

\bibitem{FP}
L.~D.~Faddeev and V.~N.~Popov,
``Feynman diagrams for the Yang-Mills field,''
Phys. Lett. B \textbf{25}, 29 (1967).
  
\bibitem{FT1982} 
  E.~S.~Fradkin and A.~A.~Tseytlin,
  ``Asymptotic freedom in extended conformal supergravities,''
  Phys.\ Lett.\ B {\bf 110}, 117 (1982);
  ``One-loop beta function in conformal supergravities,''
  Nucl.\ Phys.\ B {\bf 203}, 157 (1982).

  
\bibitem{Paneitz}
  S.~M.~Paneitz,
 ``A quartic conformally covariant differential operator for 
 arbitrary pseudo-Riemannian manifolds,'' MIT preprint, March 1983; 
 published posthumously in:  SIGMA {\bf 4},  036 (2008)
  [arXiv:0803.4331 [math.DG]].

\bibitem{Riegert}
  R.~J.~Riegert,
  ``A non-local action for the trace anomaly,''
  Phys.\ Lett.\ B {\bf 134}, 56 (1984).

\bibitem{Babaei-Aghbolagh:2022uij}
H.~Babaei-Aghbolagh, K.~B.~Velni, D.~M.~Yekta and H.~Mohammadzadeh,
``Emergence of non-linear electrodynamic theories from $T\bar T$-like deformations,''
Phys. Lett. B \textbf{829}, 137079 (2022)
[arXiv:2202.11156 [hep-th]].

\bibitem{FST-M1}
C.~Ferko, L.~Smith and G.~Tartaglino-Mazzucchelli,
``On current-squared flows and ModMax theories,''
SciPost Phys. \textbf{13}, no.2, 012 (2022) 
[arXiv:2203.01085 [hep-th]].

\bibitem{FST-M2}
C.~Ferko, L.~Smith and G.~Tartaglino-Mazzucchelli,
``Stress tensor flows, birefringence in non-linear electrodynamics and supersymmetry,''
SciPost Phys. \textbf{15}, no.5, 198 (2023) 
[arXiv:2301.10411 [hep-th]].

\bibitem{Ferko:2023wyi}
C.~Ferko, S.~M.~Kuzenko, L.~Smith and G.~Tartaglino-Mazzucchelli,
``Duality-invariant nonlinear electrodynamics and stress tensor flows,''
Phys. Rev. D \textbf{108}, no.10, 106021 (2023)
[arXiv:2309.04253 [hep-th]].

\bibitem{KMcC}
S.~M.~Kuzenko and S.~A.~McCarthy,
``Nonlinear self-duality and supergravity,''
JHEP {\bf 0302}, 038 (2003)
[hep-th/0212039].  

\bibitem{KMcC2}
S.~M.~Kuzenko and S.~A.~McCarthy,
``On the component structure of N=1 supersymmetric nonlinear electrodynamics,''
JHEP \textbf{05}, 012 (2005)
[arXiv:hep-th/0501172 [hep-th]].


\bibitem{HT}
P.~S.~Howe and R.~W.~Tucker,
``Scale invariance in superspace,''
Phys.\ Lett.\ B {\bf 80}, 138 (1978).


\bibitem{K13} 
S.~M.~Kuzenko,
``Duality rotations in supersymmetric nonlinear electrodynamics revisited,''
JHEP {\bf 1303}, 153 (2013)
[arXiv:1301.5194 [hep-th]].

\bibitem{ILZ}
E.~Ivanov, O.~Lechtenfeld and B.~Zupnik,
``Auxiliary superfields in N=1 supersymmetric self-dual electrodynamics,''
JHEP \textbf{05}, 133 (2013)
[arXiv:1303.5962 [hep-th]].

\bibitem{Kuzenko:2021qcx}
S.~M.~Kuzenko and E.~S.~N.~Raptakis,
``Duality-invariant superconformal higher-spin models,''
Phys. Rev. D \textbf{104}, no.12, 125003 (2021)
[arXiv:2107.02001 [hep-th]].

\bibitem{Kuzenko:2019vaw}
S.~M.~Kuzenko,
``Superconformal vector multiplet self-couplings and generalised Fayet-Iliopoulos terms,''
Phys. Lett. B \textbf{795}, 37-41 (2019)
[arXiv:1904.05201 [hep-th]].

\bibitem{Kuzenko:2023igt}
S.~M.~Kuzenko and J.~C.~Stirling,
``New duality-invariant models for nonlinear supersymmetric electrodynamics,''
JHEP \textbf{12}, 041 (2023)
[arXiv:2308.07113 [hep-th]].

\bibitem{ButterK13}
D.~Butter and S.~M.~Kuzenko,
``Nonlocal action for the super-Weyl anomalies: A new representation,''
JHEP {\bf 1309}  (2013) 067 
[arXiv:1307.1290 [hep-th]].


\bibitem{BdeWKL}
D.~Butter, B.~de Wit, S.~M.~Kuzenko and I.~Lodato,
``New higher-derivative invariants in N=2 supergravity and the Gauss-Bonnet term,''
JHEP {\bf 1312},  062  (2013) 
[arXiv:1307.6546 [hep-th]].

\bibitem{GKP}
D.~T.~Grasso, S.~M.~Kuzenko and J.~R.~Pinelli, {\it Work in progress}.

\bibitem{Bandos:2020hgy}
I.~Bandos, K.~Lechner, D.~Sorokin and P.~K.~Townsend,
``On p-form gauge theories and their conformal limits,''
JHEP \textbf{03}, 022 (2021)
[arXiv:2012.09286 [hep-th]].


\bibitem{Ferko:2024zth}
C.~Ferko, S.~M.~Kuzenko, K.~Lechner, D.~P.~Sorokin and G.~Tartaglino-Mazzucchelli,
``Interacting chiral form field theories and $T\overline T$-like flows in six and higher dimensions,''
[arXiv:2402.06947 [hep-th]].


\bibitem{Pasti:1996vs}
P.~Pasti, D.~P.~Sorokin and M.~Tonin,
``On Lorentz invariant actions for chiral p forms,''
Phys. Rev. D \textbf{55}, 6292-6298 (1997)
[arXiv:hep-th/9611100 [hep-th]].

\bibitem{Pasti:1997gx}
P.~Pasti, D.~P.~Sorokin and M.~Tonin,
``Covariant action for a D = 11 five-brane with the chiral field,''
Phys. Lett. B \textbf{398}, 41-46 (1997)
[arXiv:hep-th/9701037 [hep-th]].

\bibitem{Ivanov:2014nya}
E.~A.~Ivanov, A.~J.~Nurmagambetov and B.~M.~Zupnik,
``Unifying the PST and the auxiliary tensor field formulations of 4D self-duality,''
Phys. Lett. B \textbf{731}, 298-301 (2014)
[arXiv:1401.7834 [hep-th]].

\bibitem{Mkrtchyan:2019opf}
K.~Mkrtchyan,
``On covariant actions for chiral $p-$forms,''
JHEP \textbf{12}, 076 (2019)
[arXiv:1908.01789 [hep-th]].

\bibitem{Bansal:2021bis}
S.~Bansal, O.~Evnin and K.~Mkrtchyan,
``Polynomial duality-symmetric Lagrangians for free p-forms,''
Eur. Phys. J. C \textbf{81}, no.3, 257 (2021)
[arXiv:2101.02350 [hep-th]].

\bibitem{Avetisyan:2022zza}
Z.~Avetisyan, O.~Evnin and K.~Mkrtchyan,
``Nonlinear (chiral) p-form electrodynamics,''
JHEP \textbf{08}, 112 (2022)
[arXiv:2205.02522 [hep-th]].



\end{thebibliography}
\end{document}